\documentclass[prb,showpacs,amsmath,amssymb,superscriptaddress,twocolumn,floatfix]{revtex4}

\usepackage{graphicx}
\usepackage{dcolumn}
\usepackage{bm}
\usepackage{epsfig}

\begin{document}


\title{Modeling of diffusion of injected electron spins in spin-orbit coupled microchannels}
\author{Liviu P. Z\^ arbo}
\affiliation{Institute of Physics ASCR, v.v.i., Cukrovarnick\'a 10, 162 53 Praha 6, Czech Republic}
\author{Jairo Sinova}
\affiliation{Department of Physics, Texas A\& M
University, College Station, Texas 77843-4242, USA}
\affiliation{Institute of Physics ASCR, v.v.i., Cukrovarnick\'a 10, 162 53 Praha 6, Czech Republic}
\author{I. Knezevic}
\affiliation{Department of Electrical and Computer Engineering
University of Wisconsin-Madison
Madison, Wisconsin 53706, USA}
\author{J.~Wunderlich}
\affiliation{Hitachi Cambridge Laboratory, Cambridge CB3 0HE, United Kingdom}
\affiliation{Institute of Physics ASCR, v.v.i., Cukrovarnick\'a 10, 162 53
Praha 6, Czech Republic}
\author{T.~Jungwirth}
\affiliation{Institute of Physics ASCR, v.v.i., Cukrovarnick\'a 10, 162 53 Praha 6, Czech Republic}
\affiliation{School of Physics and
Astronomy, University of Nottingham, Nottingham NG7 2RD, United Kingdom}

\begin{abstract}
We report on a theoretical study of spin dynamics of an ensemble of  
spin-polarized electrons injected in a diffusive microchannel
with 
linear Rashba and Dresselhaus spin-orbit coupling. 
 We explore the dependence of the spin-precession 
and spin-diffusion lengths on  the 
strengths of spin-orbit interaction and external magnetic fields, microchannel width, 
and orientation. Our results are based on numerical Monte Carlo simulations and 
on approximate analytical formulas, both treating the spin dynamics 
quantum-mechanically.  We conclude that spin-diffusion lengths comparable or 
larger than the precession-length occur i) in the vicinity of the persistent 
spin helix regime for arbitrary channel  width, and ii) in channels of similar
 or smaller width than the precession length, independent of the ratio of 
Rashba and Dresselhaus fields. For similar strengths of the Rashba and 
Dresselhaus fields, the steady-state spin-density oscillates or remains 
constant along the channel  for channels parallel to the in-plane 
diagonal 
crystal directions. 
An oscillatory spin-polarization pattern tilted by 
45$^{\circ}$ with respect to the channel axis is predicted for channels 
along the main cubic crystal directions.
For typical experimental system 
parameters, magnetic fields of the order of tesla are required to 
affect the spin-diffusion and spin-precession lengths.
\end{abstract}

\maketitle


\section{Introduction}
\label{sec:introd}


Spin-orbit (SO) coupling in vacuum is a relativistic effect in which the magnetic
moment of a moving electron couples to an external electric field.
The effect can be explained by recalling that the moving 
magnetic moment is seen in the laboratory frame as both magnetic and electric
dipole moment and the electric dipole component couples to the external electric
field. The correct magnitude of the SO coupling 
term can be derived using the Dirac equation
for the moving particle. Owing to the band structure, the SO
coupling for electrons in solids can be enhanced by orders of magnitude
with respect to the value computed in vacuum.
This makes the SO coupling-based effects experimentally accessible
and enables the use of SO coupling as a tool for purely electrical generation
and manipulation of spins in devices.\cite{vZuti'c2004,Fabian2007,Sinova2008,Wu2010}

The prototype spintronic device using SO coupling as a spin
control tool is the Datta-Das transistor.\cite{Datta1990} 
It consists of a SO coupled channel connected to spin 
polarized source and drain electrodes. Inside the channel, the electron 
undergoes coherent spin rotations under the influence of the SO
field which can be tuned electrically by an external gate. However, the simplicity of the
Datta-Das concept is deceptive 
whenever the channel is not one-dimensional. 
The main problem concerning spin
transport in the channel of a Datta-Das device is that
on one hand, the SO coupling strength in the channel
has to be large enough to enable control of the electron spin.
On the other hand, however, a large SO coupling can lead to a faster spin
relaxation via 
 D'yakonov-Perel\cite{D'yakonov1971a} mechanism than the electron 
dwell time in the channel if the channel is not 
one-dimensional or ballistic. 
To overcome the difficulty, it was 
 proposed\cite{Schliemann2003a,Cartoixa2003}
to exploit the symmetry arising from the interplay
of  Rashba\cite{Bychkov1984} and Dresselhaus\cite{Dresselhaus1955}
 SO fields in the two-dimensional electron gas (2DEG) formed in
semiconductor heterostructures. This proposal has opened a 
way to the Datta-Das transistor
operating in a non-ballistic regime.\cite{Cartoixa2003,Schliemann2003a,Hall2003}

SO fields in the 2DEG act as 
 momentum dependent magnetic fields that couple to the electronic
magnetic moment. Impurities, phonons, or crystalline defects can 
scatter the electrons which changes their momenta and, therefore, changes the
effective SO-induced magnetic field acting on the electron spin. 
Individual electron spins in the channel
acquire different phases with respect to each other, resulting in
the relaxation of the total spin. This is the qualitative picture of
the D'yakonov-Perel relaxation. 

The idea behind the non-ballistic
Datta-Das spin transistor is that one could tune the Rashba 
and Dresselhaus SO coupling strengths to be equal, e.g. 
via gate voltage.\cite{Nitta1997}
In this case, the orientation of the total Rashba-Dresselhaus SO
 field is independent of momentum and is parallel to one of the 
in-plane diagonal axes (that can be either $[1\overline{1}0]$ or $[110]$  
depending on the relative sign of the Rashba and Dresselhaus fields)
in the (001)-plane of the 2DEG in a cubic 
semiconductor. The amplitude of the SO field depends only on the 
momentum component perpendicular to the direction of the SO 
field.\cite{Schliemann2003a,Bernevig2006c} This can lead to
a path independent spin precession of individual electron
spins, and thus to a suppression of the spin relaxation, 
in 2DEG channels oriented perpendicular to the linear 
Rashba-Dresselhaus SO field. 
Moreover, the Hamiltonian exhibits the U(1) symmetry 
which means that an in-plane spin state parallel  to this SO field
direction is infinitely long lived. 
This state will be dephased if
the cubic Dresselhaus term is present in the 
system.\cite{Stanescu2007,Cheng2006,Cheng2007a}
Randomness in the SO coupling induced by remote impurities would 
cause additional spin relaxation.\cite{Sherman2005} 
Nevertheless, infinite spin lifetimes are still
possible in SO coupled 2DEGs if the spatially varying SO field can be 
described as a pure gauge and, thus,
 removed by a gauge transformation.\cite{Tokatly2010}

Furthermore, it was shown\cite{Bernevig2006c} that the 
many-electron system whose individual particles are described 
by the above U(1) symmetric single-particle Hamiltonian 
displays a  SU(2) symmetry which is robust against both spin-independent
disorder and electron-electron interactions. Owing to this symmetry, a collective
spin state excited at a certain wave vector would have an infinite 
lifetime. Such a state is called the persistent spin helix\cite{Bernevig2006c} (PSH)
and it has already been observed in transient spin grating
experiments.\cite{Weber2007,Koralek2009} 

In  another recent experiment,\cite{Wunderlich2009} spin polarized current
passing through a micrometer-size 2DEG channel has been detected by measuring the 
SO coupling-induced Hall signal. This is called the spin
injection Hall effect (SIHE). The fact that the SIHE 
observed in a diffusive channel is robust against disorder 
and temperature effects and that the estimated
Rashba and Dresselhaus SO couplings are similar in the 2DEG 
system employed in the experiment leads 
to the question whether the PSH physics is relevant to this 
transport experiment.

In this paper, we investigate theoretically   
spin dynamics of electrons in the 2DEG channel in the PSH 
regime as well as in regimes of different Rashba and Dresselhaus 
SO field strengths. In the context of the above SIHE experiment 
we point out in particular that 
spin diffusion lengths comparable to the spin precession length
occur in channels whose widths are smaller or comparable to the spin 
precession length, regardless of the ratio between the 
Rashba and Dresselhaus SO coupling strengths. This is one of the 
several conclusions of the calculations presented below which
 consider the dependence of the spin diffusion characteristics 
on experimentally relevant system parameters such as the strengths of 
SO and external magnetic fields, microchannel width, and orientation.

In our calculations we employ the non-interacting electron 
approximation\cite{Kiselev2000,Ohno2008}  and consider the 
diffusive regime in which the SO
splitting is much smaller than the energy level broadening
due to disorder scattering, $\Delta_{\rm SO}\ll \hbar/\tau$. 
In this approach, momentum and position of electrons can be treated as 
classical variables.
 We emphasize that the direct correspondence mentioned above 
between the suppressed spin relaxation in the single-particle 
transport problem and the collective PSH state  is valid in 
this diffusive regime. Here the group velocity of an electron 
in the Rashba-Dresselhaus 2DEG can be approximated by its 
momentum divided by the mass. The spin precession angle of 
such a particle depends only on the 
distance traveled along the direction perpendicular to the 
SO  field.\cite{Bernevig2006c}
The resulting spin density pattern of
an ensemble of injected electron spins then coincides with the spin density 
pattern of the PSH spin wave. 
The expression of velocity of SO-coupled electrons contains terms proportional 
to SO coupling strength. For example, the velocity 
along $[1\overline{1}0]$-direction of Rashba and Dresselhaus 
SO coupled electrons in the PSH regime ($\alpha=-\beta$) is
$\mathbf{v_{1\overline{1}0}}=\hbar\mathbf{k_{1\overline{1}0}}/m^*\pm 2\beta/\hbar$, where 
$\alpha=-\beta$ are the Rashba and Dresselhaus SO coupling strengths. 
This means that in the opposite limit of strong SO coupling and weak disorder, the velocity and 
momentum are not simply proportional to each other and the 
direct link is lost between the one-particle and collective 
physics in the regime of equal or similar Rashba and Dresselhaus field strengths. 

The paper is organized as follows:
In Sec.~\ref{sec:theory} we introduce our method and discuss
our approximations.
In Sec.~\ref{sec:mc} we outline the features of the 
Monte Carlo method we use in our simulations. 
In  Sec.~\ref{sec:so_psh} we discuss the single particle evolution of
quantum spin in SO coupled 2DEG and its dependence on crystalline direction
of propagation, external magnetic field and interplay of Rashba and 
Dresselhaus SO couplings. 
In Sec.~\ref{sec:results} 
we show how the steady state spin density distribution of  an ensemble of electrons in
the channel 
is affected by varying the above parameters. Sec.~\ref{sec:concls} 
gives the main conclusions of our work.

\begin{figure}[t]
  \centering
  \includegraphics[scale=0.34]{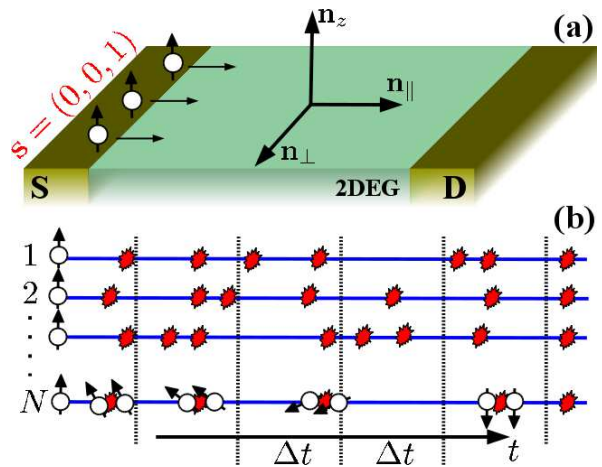}
  \caption[device]{(Color online) (a) Representation of the our model device. 
Spin-$\uparrow$ polarized particles are injected from the source electrode in a Rashba and
Dresselhaus SO coupled channel. 
(b) Schematic depiction of the EMC method. The time evolution of 
each particle belonging to the ensemble is sampled at equal intervals
$\Delta t$ called subhistories. The particle
spin precesses in the SO field during the free flight time, but is 
unaffected by collisions.
 }
  \label{fig:sample}
\end{figure}

\section{Theoretical model}
\label{sec:theory}

We are interested in the spin dynamics in the 2DEG channel
of an experimentally relevant
spintronic model device  which is schematically depicted in 
Fig.~\ref{fig:sample}(a). 
The typical device is of a few micrometers in size and this
is considerably larger than the Fermi wave length in the 2DEG
channel. 
 It means that the quantum interference effects on
the orbital motion of electrons can be neglected. In other
words, it is sufficient to solve Boltzmann transport equation (BTE)
for this system, rather than use a fully quantum mechanical treatment
such as Keldysh formalism. On the other 
hand, we cannot neglect  quantum mechanics of the spin
dynamics since the typical spin precession length in experiments\cite{Wunderlich2009}
ranges from a few hundreds of nanometers to a few microns.
Therefore,
we employ  the ensemble 
Monte Carlo\cite{Hockney1981,Jacoboni1983a,Fischetti1988,Fischetti1993} (EMC) 
method which is a well established tool in semiconductor
device simulations and can be extended to include spin coherence\cite{Saikin2003} 
in a micrometer size device. 
The EMC method offers a way to solve BTE that
is beyond the reach of drift-diffusion models. 
Drift-diffusion models must rely on various approximations
in order to avoid the tremendous mathematical difficulties 
arising in BTE. Treatment of nonlinear terms, inclusion of different
scattering mechanisms, or of dissipation effects
require drastic approximations so that the result of the
drift-diffusion calculation might not reflect anymore the 
features of the theoretical model, but rather those of the
mathematical approximations. By contrast, including 
disorder, dissipation, temperature, transient or nonlinear effects
in EMC simulations is straightforward and does not 
require  further approximations. In fact, state-of-the-art
EMC simulations are often used to test the validity of
the drift-diffusion models. Inclusion of a various range of
effects in the EMC is done without significant changes in the
computational complexity, thus making it more suitable for
device simulation than other powerful quantum mechanical techniques
such as nonequilibrium Green function (NEGF) method.
For example, including dissipative effects  in
EMC has only a minor impact on the calculation complexity, while in the case 
of NEGF method it can reduce the size of computationally accessible systems
from a few hundreds to a few tens of nanometers.

In our calculations, we make the following approximations: 
(i) Electron orbital degrees of freedom are 
described by classical momentum and position and the spin degree of freedom by 
quantum-mechanical spin density matrix.
This semiclassical approximation is justified by the diffusive regime we consider.
In this regime we can approximate the electron velocity by 
momentum divided by mass.
(ii) Interactions between electrons are neglected.
(iii) We consider only short range impurity scattering. 
(iv) Temperature enters our simulations only through the Fermi-Dirac
distribution function.
(v) For simplicity, we neglect the electrostatics of the channel.
This is justified since we are primarily interested in the spin dynamics
 of electrons. A small electric field present in the channel is not
expected to have an important influence on the spin precession
pattern of the electronic system. 

In the next two subsections we briefly outline the spin dependent EMC method
and analyze the motion of a single particle in the SO field.

\subsection{Spin Dependent Monte Carlo}
\label{sec:mc}

Electrons in the channel, shown in Fig.~\ref{fig:sample}(a), can be modeled 
as an ensemble of $N$ noninteracting particles. In the EMC method, we track the
individual motion of each particle in the ensemble and we use the data to 
calculate an approximate particle distribution in phase space. As shown in
Fig.~\ref{fig:sample}(b), we divide the time of simulation in small time intervals
$\Delta t$
called subhistories. During the subhistory, a particle moving in 
electromagnetic and SO
fields in the channel can be randomly scattered by impurities and, 
in general, also by phonons or other scattering mechanisms that
are present in the channel.
 The time between collisions, called the ``free flight time'',
is randomly generated and depends on the scattering rate corresponding 
to each type of collision.\cite{Jacoboni1983a} Fig.~\ref{fig:sample}(b) 
gives an intuitive picture of the time evolution of the ensemble of electrons.
The semiclassical particle is described
by its position $\mathbf{r}(t)$ and  momentum $\mathbf{k}(t)$.
 As it was recently 
shown,\cite{Kiselev2000,Saikin2003} we can treat 
spin-dependent phenomena if, in addition to the semiclassical 
variables, we consider that each particle is described by a $2\times 2$
spin density matrix $\hat{\rho}(t)$. The spin polarization vector
is then given by $\mathbf{s}={\rm Tr}[\hat{\rho}{\bm \sigma}]$. The propagation of 
a particle
during the free flight is described by the equations of motion
for its attached dynamical variables
\begin{subequations}
  \begin{equation}
    \label{eq:r}
    m^*\frac{d^2\mathbf{r}}{dt^2}=-e[\mathbf{E}+\mathbf{v_k}\times \mathbf{B}],
  \end{equation}
  \begin{equation}
    \label{eq:p}
    \mathbf{v_k}=\frac{1}{\hbar}\nabla_\mathbf{k}E_\mathbf{k} \approx  
    \frac{\hbar\mathbf{k}}{m^*},
  \end{equation}
  \begin{equation}
    \label{eq:spin_evol}
    \hat{\rho}(t+\delta t)=e^{-\frac{i}{\hbar}\hat{H}_{\rm spin}(\mathbf{k})\delta t}
    \hat{\rho}(t) e^{\frac{i}{\hbar}\hat{H}_{\rm spin}(\mathbf{k})\delta t},
  \end{equation}
\end{subequations}
which must be integrated together to find the particle time evolution
during free flights. 
Here, $\mathbf{v_k}$ is the particle velocity, $\mathbf{E}$ and $\mathbf{B}$
are the electric and magnetic fields, $m^*$ is the effective mass of the 
particle, and  $E_\mathbf{k}$ is the electronic band dispersion in the 2DEG.
Note that the approximation in Eq.~(\ref{eq:p}) means that we neglect 
any influence of the SO coupling on the trajectory of the semiclassical 
particle,\cite{Sinitsyn2008}.
While equations~(\ref{eq:r}) and~(\ref{eq:p}) describe a semiclassical 
electron propagating in a solid,  Eq.~(\ref{eq:spin_evol})
describes the quantum mechanical evolution of its spin 
during short time $\delta t$ which is 
controlled by the spin-dependent part of the 
Hamiltonian, $\hat{H}_{\rm spin}$. This Hamiltonian includes the internal 
SO field and the external magnetic field, 
$\hat{H}_{\rm spin}=\hat{H}_{\rm SO}+\hat{H}_{\rm Z}$.

At the end of each subhistory we calculate the ensemble averaged 
quantities of interest such as currents, charge and spin densities.
After the simulation converged and the system is in steady state
we can use the subsequent subhistories to compute time averaged
values for the physical quantities. In our case, the system is
in steady state when the flux of electrons through the 
drain electrode becomes constant.

\subsection{Spin Dynamics in Rashba-Dresselhaus Field}
\label{sec:so_psh}

The electron gas in the heterostructure can be modeled by the Rashba and Dresselhaus SO
coupled Hamiltonian
\begin{equation}
  \label{eq:rash-dres}
  \hat{H}=\frac{\hat{\mathbf{p}}^2}{2m^*}+
  \frac{\alpha}{\hbar}
  \left( \hat{p}_y\sigma_x- \hat{p}_x\sigma_y \right)
  +
  \frac{\beta}{\hbar}
  \left( \hat{p}_x\sigma_x- \hat{p}_y\sigma_y \right)\,.
\end{equation}
Here $\beta$ is the Dresselhaus SO coupling which, for simplicity, is kept constant
in our simulations, $\alpha$ is the experimentally 
adjustable\cite{Nitta1997,Lechner2009} Rashba parameter, $m^*$ is the  effective mass 
of the 2DEG, and $\sigma_x$, $\sigma_y$, and $\sigma_z$ are the Pauli
matrices. The crystalline axes labeled $x$, $y$, and $z$ correspond to the
$[100]$, $[010]$, and $[001]$ directions, respectively, and the 2DEG lies in the $xy$-plane.
We are ignoring the cubic Dresselhaus terms since the linear terms are dominant
for not too high carrier concentrations.
The effect of the magnetic field is included
by making the substitution $\mathbf{p}\rightarrow \mathbf{p}-e\mathbf{A}$
in the Hamiltonian~(\ref{eq:rash-dres}) and by adding
the Zeeman term,
\begin{eqnarray}
  \label{eq:zeeman}
  \hat{H}_{\rm Z} & = 
  &
 - \frac{1}{2}g\mu_{\rm B}\mathbf{B}\cdot{\bm \sigma} \\
  & = &   -  \frac{1}{2}g\mu_{\rm B}
  (B_\parallel\sigma_\parallel+B_\perp\sigma_\perp+B_z\sigma_z ),
  \nonumber
\end{eqnarray}
where $\mathbf{B}$ is the magnetic field strength, $\mathbf{A}$ is the corresponding
vector potential, $g$  is the $g$-factor, and $\mu_{\rm B}$ is the Bohr magneton. 
We wrote the Zeeman Hamiltonian in terms of the magnetic field components 
$B_\parallel$ which is parallel to the transport direction in the channel, 
$B_\perp$ which is the in-plane magnetic field component perpendicular to the
current direction, and $B_z$ which is the out-of-plane component of the magnetic
field. The unit vectors corresponding to in-plane axes parallel and 
perpendicular to the transport direction are labeled by
$\mathbf{n}_\parallel=(a,b,0)$ and $\mathbf{n}_\perp=(b,-a,0)$ while $\mathbf{n}_z=(0,0,1)$
corresponds to the $z$-axis. For example, if the electronic transport is along 
$[1\overline{1}0]$-axis, $a=1/\sqrt{2}$ and  $b=-1/\sqrt{2}$. Using this notation
we can express the spin matrices as 
$\sigma_\parallel=\mathbf{n}_\parallel {\bm \sigma}=a\sigma_x+b\sigma_y$ and
$\sigma_\perp=\mathbf{n}_\perp {\bm \sigma} = b\sigma_x-a\sigma_y$.
Note that the above Hamiltonian~(\ref{eq:zeeman}) does not take into account
the change in the effective mass or $g$-factor in the 2DEG as a result
of applying magnetic field.\cite{Winkler2003}
   
In what follows, we derive the spin precession length of an electron propagating 
in a straight line
along an arbitrary direction in the SO coupled 2DEG. As in the spin-dependent EMC
 approach described in the previous section, the electron is a point particle
whose spin rotates coherently under the influence of a weak SO field 
and the applied external magnetic field.
 We consider a spin-$\uparrow$  electron (spin parallel to +$\hat z$-direction) injected 
along an arbitrary direction $\mathbf{n}_\parallel=(a,b,0)$ and subject to 
both SO and magnetic fields. The electron spin is described by the spin
density matrix $\rho=\frac{1}{2}({\bf I}_2+\mathbf{s}{\bm \sigma})$, where 
$\mathbf{s}=(s_x,s_y,s_z)$ is the spin polarization vector and 
${\bf I}_2$ is the $2\times 2$ identity matrix.
Initially,
the electron has spin-$\uparrow$, so its polarization vector is $\mathbf{s}=(0,0,1)$
and the spin density matrix is $\rho_0=\frac{1}{2}({\bf I}_2+\sigma_z)$.

Next, we rewrite the SO part of the Hamiltonian~(\ref{eq:rash-dres}) 
as 
\begin{equation}
  \label{eq:hso}
  \hat{H}_{\rm SO}=\Omega_x\sigma_x+\Omega_y\sigma_y+\Omega_z\sigma_z,
\end{equation}
with $\Omega_x=\alpha k_y+\beta k_x$, 
$\Omega_x=(\alpha k_x+\beta k_y)$ and 
$\Omega_z=0$. We label the unit vector parallel to 
${\bm \Omega}=(\Omega_x,\Omega_y,\Omega_z)$ 
by $\mathbf{h}=(h_x,h_y,h_z)$.
After a short time step $\delta t$ during which the momentum
is considered constant, we obtain with the aid of 
Eq.~(\ref{eq:spin_evol})
\begin{equation}
  \label{eq:rhodt}
  \rho(\delta t)=\frac{1}{2}{\bf I}_2+\frac{1}{2}\cos(2P\delta t)\sigma_z-
  \frac{1}{2}\sin(2P\delta t)(h_x\sigma_y-h_y\sigma_x),
\end{equation}
where 
$P=\frac{1}{\hbar}\sqrt{\Omega_x^2+\Omega_y^2}$. In order to study the single
particle spin precession we consider that the 
electron momentum
along the transport direction is constant. Such assumption is true
as long as there is no transverse external electric field and no out-of-plane
magnetic field. The condition that the spin flips during the motion 
of the electron is $2P t^{\uparrow\rightarrow\downarrow}=\pi$, as seen from 
Eq.~(\ref{eq:rhodt}). The spin precession length along the
transport direction $\mathbf{n}_\parallel=(a,b,0)$ is computed as
$L_{\rm ab0}^{\uparrow \rightarrow \downarrow}=\mathbf{v_k}t^{\uparrow \rightarrow \downarrow}$.
(Recall that $\mathbf{v_k}\approx \hbar \mathbf{k}/m^*$ in the diffusive, weak SO 
coupling regime.) Considering the effects of
the in-plane magnetic field, we obtain for the spin precession length
\begin{widetext}
\begin{equation}
  \label{eq:lso}
  L_{\rm ab0}^{\uparrow \rightarrow \downarrow}=
  \frac{\pi\hbar^2}
  {2m^*
    \sqrt{
      \left(\alpha b+\beta a+\displaystyle \frac{1}{2}g\mu_{\rm B}
        \displaystyle \frac{B_\parallel a+B_\perp b}{k}
      \right)^2
      +
      \left(-\alpha a-\beta b+\displaystyle \frac{1}{2}g\mu_{\rm B}
        \displaystyle \frac{B_\parallel b+B_\perp a}{k}
      \right)^2
    }
  }.
\end{equation}
\end{widetext}

By applying the spin precession length formula~(\ref{eq:lso}) 
we can gain an intuitive understanding of the spin dynamics in the SO coupled
heterostructure. Of particular interest is the case of $\alpha =-\beta$ 
and $[1\overline{1}0]$ channel orientation (or $\alpha =\beta$ and $[110]$-orientation)
in which the PSH symmetry is present. In this case,
the spin precession depends only on the 
distance traveled by electrons along the channel.
From Eq.~(\ref{eq:lso}) we can immediately see that for $\alpha =-\beta$ and electron 
propagating along the  $[1\overline{1}0]$-direction, the spin precession
length is the shortest while for spin propagating along the 
$[110]$-direction it is infinite.

\section{Discussion of the EMC simulations}
\label{sec:results}
We now employ the spin-dependent EMC method outlined in Sec~\ref{sec:mc} to numerically
simulate spin dynamics in the microchannel illustrated in Fig.~\ref{fig:sample}(a).
Spin-$\uparrow$ electrons are injected from the source electrode
and propagate in the SO field of the disordered 2DEG.
Electrons that reach
the microchannel edge  are reflected back with unchanged spin. An electron
that re-enters the source or exits the drain is erased and replaced 
by a new spin-$\uparrow$ electron
injected from the source. 

We choose the 2DEG parameters that correspond to the GaAs
2DEG of Ref.~[\onlinecite{Wunderlich2009}]. Our channel length is 
$L\approx 3\,{\rm \mu m}$ and the width will take values both 
smaller and larger than the spin precession length which is of the 
order of a few hundred nanometers.
Temperature of the electron ensemble in all simulations
is 300~K. Electron-phonon scattering is neglected.
Disorder in the system is due to randomly placed 
point-like spinless impurities. 
The electron-mean-free path is 26~nm. 
The electronic density of the 2DEG system is $n_e=2.5\times 10^{12}\,{\rm cm^{-2}}$.
The corresponding Fermi wave length is $\lambda_F\approx 8$nm.
The Dresselhaus spin-orbit coupling is kept constant during simulations at
 $\beta=-2.0\times 10^{-12}\,{\rm eV \cdot m}$. 

\begin{figure}[t]
  \centering
  \includegraphics[scale=0.34]{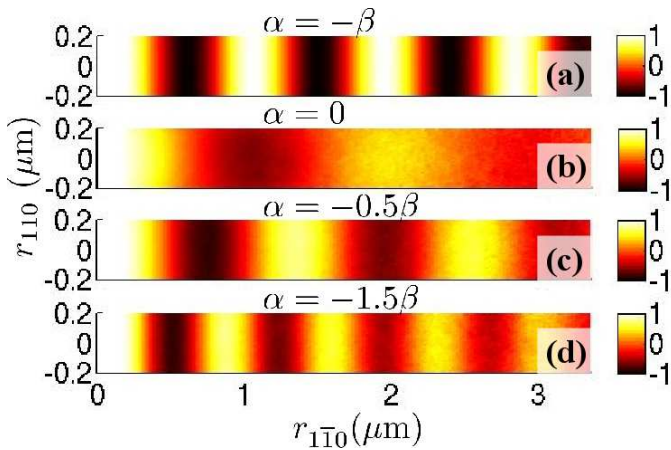}  
  \caption[rash-dress]{(Color online) Spin density distribution 
    $\left< S_z(\mathbf{r}) \right>$ in a 
    $[1\overline{1}0]$-oriented 2DEG channel with fixed Dresselhaus 
    SO coupling $\beta=-2.0\times 10^{-12}{\rm eV \cdot m}$ for
    different values of Rashba SO coupling
    (a) $\alpha= -\beta $, (b) $\alpha =0$, (c) $\alpha= -0.5 \beta $
    and (d) $\alpha= -1.5\beta $. The light color
    refers to spin-$\uparrow$. The distance between the successive
    maxima of the spin distribution corresponds to twice the spin
    precession length computed from Eq.~(\ref{eq:lso}). The spin diffusion 
    length is infinite in (a), while in all the other cases it exceeds
    the spin precession length, as expected in narrow channel.
  }
  \label{fig:fig1}
\end{figure}

The electron ensemble consists of
$N=130000$ electrons.
We run the simulation until the system reaches steady state and after that we
use the last 2000 time steps to calculate time averaged spin densities. 
The  spin densities are normalized to the number of particles present in each 
grid cell such that the spin of a cell containing all spin-$\uparrow$ 
particles is 1.
In what follows, we show how the spin density distribution
$\left< S_z(\mathbf{r}) \right>$ in the channel is affected by changes in 
the width of the channel, 
the crystalline axis along which the transport takes place, and strength of 
Rashba SO coupling and magnetic fields.

In general, the spin polarization along the channel is randomized 
due to the D'yakonov Perel 
spin dephasing mechanism which is dominant in GaAs heterostructures. 
This effect is visible, e.g., in Fig.~\ref{fig:fig1}(b). The 
spin-diffusion length depends on the parameters of the microchannel.
In the limiting case of equal Rashba and Dresselhaus SO coupling 
strengths and, e.g., $\alpha= -\beta $,  
the PSH  symmetry\cite{Bernevig2006c} 
arises in the $[1\overline{1}0]$-oriented 2DEG channel and the 
oscillatory dependence of $\left< S_z(\mathbf{r}) \right>$ on 
the coordinate along the channel is undamped, as shown in 
Fig.~\ref{fig:fig1}(a). In this case spin orientations of 
injected electrons are not randomized by scattering. The 
spin-diffusion length is infinite and the spin-precession 
length is given exactly by Eq.~\ref{eq:lso}. In 
Figs.~\ref{fig:fig1}(c) and  (d) we show that
the PSH regime is robust against sizable changes in the $\alpha/\beta $
ratio.\cite{Ohno2008}

\begin{figure}[t]
  \centering
  \includegraphics[scale=0.34]{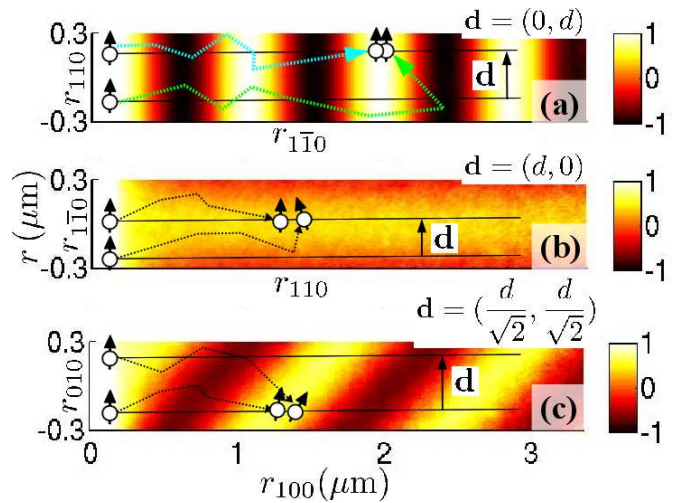}
  \caption[rash-dress]{(Color online) Spin density 
    distribution $\left< S_z(\mathbf{r}) \right>$ in the
    2DEG channel with $\alpha=- \beta $ for various orientation of
    the transport axis: (a) $[1\overline{1}0]$, (b) $[110]$ and
    (c) $[100]$. In the PSH regime, the  relative spin orientations 
    of particles injected at the source depends only on the initial
    distance between them along the $[1\overline{1}0]$-direction,
    i.e. the first component of the relative position vector of the two 
    particles $\mathbf{d}=(d_{1\overline{1}0},d_{110})$.
    In panel (a) that distance is $d_{1\overline{1}0}=0$, resulting in no spin dephasing, in
    panel (b) is $d_{1\overline{1}0}=d$ and in panel (c) it is $d_{1\overline{1}0}=d/\sqrt{2}$, where $d$ is the initial
    distance between the two particles. Therefore,the
    spin relaxation in the $[110]$ and $[100]$-oriented  channels depends on the 
    channel width.
  }
  \label{fig:diff_axes}
\end{figure}

%
%
\begin{figure}[t]
  \centering
  \includegraphics[scale=0.34]{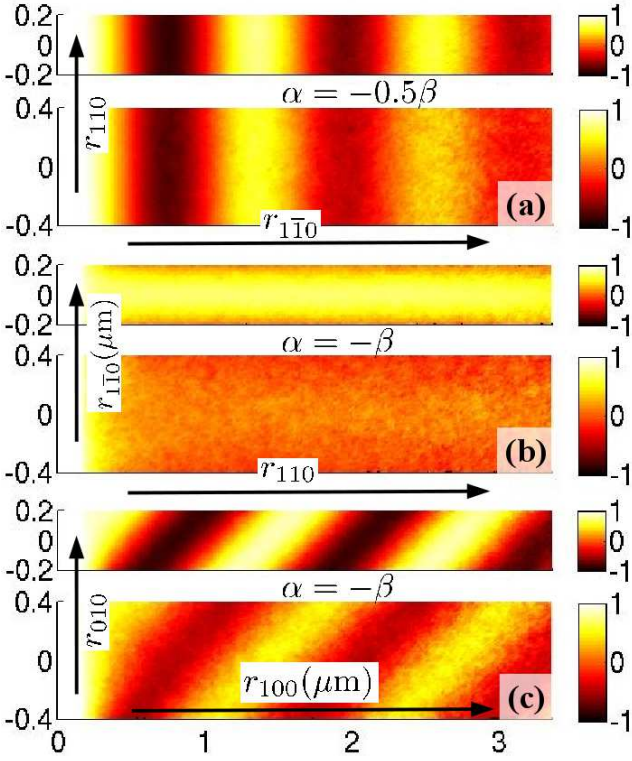}
  \caption[rash-dress]{(Color online) Spin density distribution 
    $\left< S_z(\mathbf{r}) \right>$ in the
    channel as a function of its width for
    (a) $\alpha=-0.5\beta$ and $[1\overline{1}0]$-injection,
    (b) $\alpha=-\beta$ and $[100]$-oriented channel,
    and (c) $\alpha=-\beta$ and $[110]$-injection direction. 
    As expected from theory, the larger channel widths lead 
    shorter spin diffusion lengths. 
  }
  \label{fig:widths}
\end{figure}
%

We next proceed to channels which are not 
oriented along the $[1\overline{1}0]$-direction. In Fig.~\ref{fig:diff_axes}
we compare results for the $[100]$, $[110]$-oriented channels with the
$[1\overline{1}0]$ channel,   assuming $\alpha=-\beta$.
Fig.~\ref{fig:diff_axes}(a) shows that the spin precesses fastest 
for the channel oriented along the $[1\overline{1}0]$-direction, 
while Fig.~\ref{fig:diff_axes}(b) shows no spin precession 
for the $[110]$-oriented channel, consistent with  Eq.~\ref{eq:lso}. 
The result in Fig.~\ref{fig:diff_axes}(c) for the channel oriented 
along the  $[100]$-axis is less obvious, however, 
we can still use the spin precession formula~(\ref{eq:lso}) to 
understand the 45$^\circ$ rotated oscillatory pattern for  this 
channel direction. Since the orientation of the spin depends on the distance
the particle travels  along the
$[1\overline{1}0]$-direction, we expect that a pattern formed by 
averaging the spin densities along a $[110]$-oriented line would repeat itself
with a $2L_{1\overline{1}0}^{\uparrow\rightarrow\uparrow}$ period 
along the $[1\overline{1}0]$-line. 
As sketched in Fig.~\ref{fig:diff_axes}, we can follow  
individual trajectories of two electrons injected from the source. 
%
 Let us consider two particles and connect them by the relative
position vector $\mathbf{d}$ whose length $d$ is the initial distance 
between particles. We can decompose $\mathbf{d}$ along $[1\overline{1}0]$-
and $[110]$-axes, $\mathbf{d}=(d_{1\overline{1}0},d_{110})$. 
If $\alpha=-\beta$, the difference between the spin directions 
of the two particles when they meet inside the channel is given by $d_{1\overline{1}0}$. 
The spin coherence of the electron ensemble is therefore partially 
lost because of the initial distribution of $d_{1\overline{1}0}$'s of the injected electrons. 
From this it is apparent that the spin-diffusion length scales with 
the ratio of the precession length to the channel width. 
We emphasize that all 
these arguments are independent of the mean-free-path. Indeed, 
we would  obtain the same steady state spin density distribution 
if the channel in Fig.~\ref{fig:diff_axes} were ballistic.

\begin{figure}[t]
  \centering
  \includegraphics[scale=0.25]{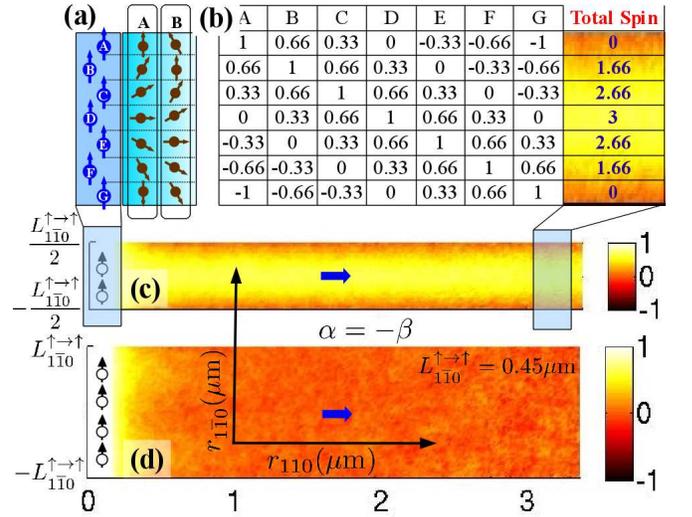}
  \caption[rash-dress]{(Color online)
    (a),(b) Pictorial explanation of the spin density patterns in 
    the $\alpha=-\beta$ regime for electrons injected along 
    a $[110]$-oriented channel of width (c) 
    $L_{1\overline{1}0}^{\uparrow\rightarrow\uparrow}$ and 
    (d) $2L_{1\overline{1}0}^{\uparrow\rightarrow\uparrow}$.
  }
  \label{fig:fig_expl}
\end{figure}

The dependence of the spin diffusion length in our ensemble of 
electrons on the channel width is further quantified in 
Fig.~\ref{fig:widths}. For the $[1\overline{1}0]$-oriented channel
the spin-diffusion length is infinite as long as $\alpha=-\beta$ 
and it decreases  with
increasing width of the channel when Rashba and Dresselhaus coupling 
strengths are not equal, as shown in Fig.~\ref{fig:widths}(a).\cite{Kiselev2000,Kettemann2007}
For  $\alpha=-\beta$, the spin-diffusion length is finite for channel orientations
different from $[1\overline{1}0]$-direction and it again decreases with increasing 
channel width. This is illustrated in Fig.~\ref{fig:widths}(b) for 
the $[110]$-oriented microchannel and in Fig.~\ref{fig:widths}(c) for the $[100]$-channel.

We now provide a more detailed understanding of the numerical 
spin-density patterns obtained by the EMC simulations, focusing 
on the $\alpha=-\beta$ case and the $[110]$-oriented microchannels. 
For fixed and equal Rashba and 
Dresselhaus coupling strengths ($\alpha=-\beta$), the spin orientation 
of an individual particle
depends only on the distance from the injection point 
along the $[1\overline{1}0]$-direction. The spin density pattern of an
ensemble of particles
starting from a given point depends only on the strength of the SO coupling. 
Our ensemble averaging procedure amounts to summing up all spin density 
patterns of particles starting from different points along the 
source-channel interface. We use this idea to 
explain the spin density distribution obtained for the $[110]$-oriented channel
of two different widths, as shown in Fig.~\ref{fig:fig_expl}.
The width of the channel in  Fig.~\ref{fig:fig_expl}(c) is equal to the 
spin precession length $L_{1\overline{1}0}^{\uparrow\rightarrow\uparrow}$. All
spins starting at the source at point A shown in Fig.~\ref{fig:fig_expl}(a)
generate a spin density pattern illustrated in the corresponding column A
in the figure. The spin density 
pattern generated by a spin starting at point B is the same, 
only shifted by the distance between A and B
along the $[1\overline{1}0]$-direction. All the other spin density patterns 
are shifted in the same manner. We can assign numbers to the projection of spin
along $z$-direction for each spin-density pattern, as shown in Fig.~\ref{fig:fig_expl}(b).
Summing up these numbers we obtain qualitatively the same transverse profile
of the spin density in the channel as in the EMC simulation. This gives an intuitive explanation 
of the reduced mean spin polarization  along the channel edges for channel width
smaller or equal to $L_{\rm SO}$ seen in Fig.~\ref{fig:fig_expl}(c). We can use the 
same procedure to explain the randomization of spins in the entire cross-section of
a wider, $[110]$-oriented channel shown in Fig.~\ref{fig:fig_expl}(d).

\begin{figure}[t]
  \centering
  \includegraphics[scale=0.34]{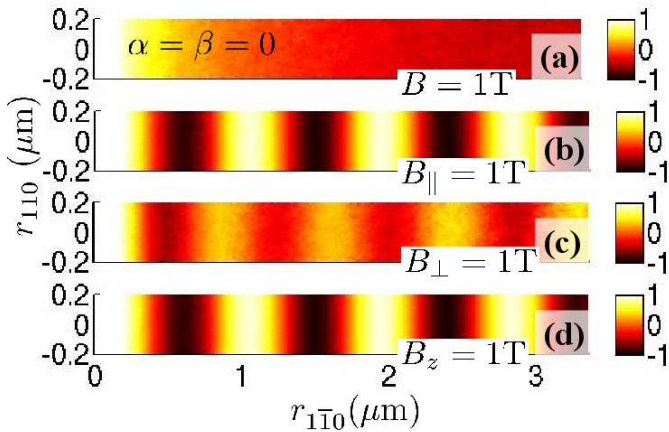}
  \caption[rash-dress]{(Color online) Spin density distribution 
    $\left< S_z(\mathbf{r}) \right>$ in magnetic 
    field for the SO coupled  2DEG channel oriented along
    $[1\overline{1}0]$-direction. The four panels show (a) Hanle effect for
    1T in-plane field and no SO coupling, spin precession for (b) $\alpha=-\beta$
    and in-plane magnetic field $B_\parallel=1$T, (c) $\alpha=-\beta$
    and in-plane magnetic field $B_\perp=1$T (d) $\alpha=-\beta$
    and out-of-plane magnetic field $B_z=1$T.
  }
  \label{fig:mag}
\end{figure}


As  apparent 
from the one particle formula~(\ref{eq:lso}), the magnetic field 
effect on the spin dynamics depends on the magnitude of the electron momentum and,
therefore, on the electron density in the channel. As shown in Fig.~\ref{fig:mag}(a),
the momentum dependence leads 
to faster randomization of the spin density distribution in the 
channel than in the momentum independent case of Dresselhaus coupled 
2DEG channel shown in  Fig.~\ref{fig:fig1}(b).
We will first analyze the effect of the in-plane magnetic 
field on spin precession. Our calculations in Fig.~\ref{fig:mag}
show that in the $\alpha=-\beta$ case and for the 
 $[1\overline{1}0]$-oriented channel, the spin density is only slightly
affected by the 
in-plane magnetic
field parallel to the channel direction even at magnitudes of the order 
of Tesla. In-plane magnetic field of the same magnitude but perpendicular to the 
channel direction has a sizable effect on the electronic spin density 
pattern. To understand the dependence on the in-plane field orientation we rewrite 
Eq.~(\ref{eq:lso}) for $\alpha=-\beta$ and for the $[1\overline{1}0]$-channel,
\begin{equation}
  \label{eq:lso2}
  L_{1\overline{1}0}^{\uparrow \rightarrow \downarrow}=
  \frac{\pi\hbar^2}
  {2m^*
    \sqrt{
      4\beta^2-\beta g\mu_{\rm B}\frac{B_\perp}{k}+
      \frac{1}{4}g^2\mu_{\rm B}^2\frac{B_\perp^2+B_\parallel^2}{k^2}
    }
  }.
\end{equation}
From Eq.~(\ref{eq:lso2}) we see that for fields $\sim 1$T, the
 magnetic field component $B_\perp$ affects the spin precession
since  the Zeeman splitting is comparable to the 
Dresselhaus spin splitting at the Fermi level. On the other hand, at higher
magnetic fields, the quadratic terms in magnetic field present in the 
denominator of Eq.~(\ref{eq:lso2})  will dominate the linear term and 
the effect of magnetic field becomes independent of its orientation. 
At these high fields, the spin-diffusion length is limited primarily 
by the magnetic field.

\section{Summary}
\label{sec:concls}

We used the spin-dependent EMC method to 
simulate spin dynamics of an ensemble of electrons
in the 2DEG channel with SO coupling. 
Our calculations were done in the diffusive regime in which 
there is a correspondence between the long spin-diffusion length of an ensemble 
of noninteracting electrons and the collective PSH state.

Our numerical simulations and qualitative analytical considerations show 
that the spin-precession pattern and the spin-diffusion length  in the 
channel with equal strengths of the Rashba and Dresselhaus SO fields depends only on 
geometric factors, i.e., on the channel orientation and width.  
The presence of magnetic field in the channel suppresses the spin-diffusion 
length. However, for the experimentally relevant 
system parameters the fields magnitude must be of the order of a
few Tesla for the effect to be observable. The presence of the PSH symmetry 
yields an oscillatory spin-density pattern with infinite spin-diffusion 
length in $[1\overline{1}0]$-oriented
channels of an arbitrary width. The spin-diffusion length is still 
comparable to the precession length
for any ratio of the Rashba and Dresselhaus fields as long as the 
channel width is comparable
or smaller than the spin-precession length. These predictions are 
independent of the scattering mean-free-path. 

\section*{Acknowledgments}

We acknowledge support  from EU Grant FP7-215368 SemiSpinNet, from 
Czech Republic Grants AV0Z10100521, KAN400100652, LC510, and 
Praemium Academiae, and from NSF-MRSEC DMR-0820414, DMR-0547875, 
and SWAN-NRI. J. S. is a Cottrell
Scholar of Research Corporation. L.P.Z. would like to thank Max Fischetti for
correspondence.


%

\end{document}